\documentclass[twocolumn,english,pra,aps,superscriptaddress,showpacs]{revtex4-1}
\usepackage[T1]{fontenc}
\usepackage[latin9]{inputenc}
\usepackage{float}
\usepackage{amsmath}
\usepackage{amssymb}
\usepackage{graphicx}

\begin{document}

\title{Implementation of a spatial two-dimensional quantum random
walk with tunable decoherence}

\author{J. Svozil\'{i}k}

\affiliation{ICFO---Institut de Ciencies Fotoniques, Mediterranean
Technology Park, 08860 Castelldefels, Barcelona, Spain}
\affiliation{Palack\'{y} University, RCPTM, Joint Laboratory of Optics, 17.listopadu 12, 771 46 Olomouc, Czech Republic}

\author{R. de J. Le\'{o}n-Montiel}

\affiliation{ICFO---Institut de Ciencies Fotoniques, Mediterranean
Technology Park, 08860 Castelldefels, Barcelona, Spain}

\author{J. P. Torres}
\affiliation{ICFO---Institut de Ciencies Fotoniques, Mediterranean
Technology Park, 08860 Castelldefels, Barcelona, Spain}
\affiliation{Department of Signal Theory and Communications, Universitat
Polit{\`{e}}cnica de Catalunya, Castelldefels, 08860 Barcelona, Spain}

\pacs{03.67.-a,03.67.Ac 03.65.Yz}

\begin{abstract}
We put forward a new, versatile and highly-scalable experimental
setup for the realization of discrete two-dimensional quantum
random walks with a single-qubit coin and tunable degree of
decoherence. The proposed scheme makes use of a small number of simple optical
components arranged in a multi-path Mach-Zehnder-like
configuration, where a weak coherent state is injected. Environmental effects
(decoherence) are generated by a spatial light modulator, which
introduces pure dephasing in the transverse spatial plane,
perpendicular to the direction of propagation of the light beam.
By controlling the characteristics of this dephasing, one can
explore a great variety of scenarios of quantum random walks: pure quantum
evolution (ballistic spread), fast fluctuating environment leading
to a diffusive classical random walk, and static disorder
resulting in the observation of Anderson localization.

\end{abstract}
\maketitle

\section{Introduction}
The underlying principles of many quantum information protocols
can be traced back to the concept of quantum random walk (QRW),
which since its first description have become a fundamental
paradigm in quantum science \cite{Douglas2009,Shenvi2003}. The
idea of QRWs was originally conceived by Aharonov \emph{et al.}
\citep{Aharonov1993} as an extension of the well-known classical
random walk (CRW) \cite{van2007stochastic}. The main
distinguishing feature of a QRW, compare to a CRW, is the possibility of
interference between the multiple paths that can be simultaneously
traversed by a quantum walker, enabling thus a faster spreading of
the uncertainty of location of the walker than in the classical case
\cite{Kempe2003,Kendon2006}.

The temporal evolution of a quantum system, such as a QRW, depends
on the presence, and specific characteristics, of the
environmental effects (decoherence) which can modify it
\cite{Schlosshauer:1105882}. In most cases, the influence of
decoherence during the evolution of a quantum walker transforms an
originally pure state into a mixed state, lowering the uncertainty about the location of the walker
as it propagates. In the limiting case, when all cross-interference terms between
different lattice sites are completely erased, the state of pure
diffusive classical propagation is reached \cite{Kendon2006}.

QRWs have been theoretically explored for the case of
one-dimensional lattices \cite{Ambainis2001,Kempe2003}, and
experimentally implemented by means of different physical
platforms, such as photon-based systems
\cite{Peruzzo2010,Broome2010,Schreiber2010,Padley2011,Sansoni2012,Schreiber2011},
optical lattices \cite{Cote2006} and waveguide arrays
\cite{Perets2008}. Also, QRWs have been implemented using trapped
ions \cite{Zahringer2010} and nuclear magnetic resonance systems
\cite{Du2003}.

Although the implementation of one-dimensional QRWs has showed to
be useful when describing several quantum information systems,
there is a great interest in expanding the concept to
multidimensional lattices. Along these lines, two-dimensional QRWs
provide a powerful tool for modeling complex quantum information
and energy transport systems
\cite{oliveira2006decoherence,schreiber20122d}. Notwithstanding,
their realization represents a challenge because of the need of a
four-level coin operation
\cite{inui2004localization,watabe2008limit,hamilton2011quantum}.
One way to overcome this drawback is to make use of different
degrees of freedom of photons, such as polarization and orbital
angular momentum, as it has been shown in \cite{schreiber20122d}.
Another approach is to mimic the two-dimensional QRWs evolution by
performing two subsequent one-dimensional QRWs
\cite{Franco2011,Franco2011A}.

Here, we make use of the latter approach to put forward an
experimental setup for the realization of  two-dimensional QRWs.
We include the environmental effects (decoherence) as pure
dephasing by means of the introduction of random phase
patterns, generated by a spatial light modulator (SLM), which
can be different from site to site ({\em spatial disorder}). By
controlling the degree of decoherence, we study the transition
from the quantum ballistic spreading to the diffusive classical
walk. Also, by adding static disorder, we show the possibility of
observing Anderson localization \citep{anderson1958absence}.

Importantly, our proposal provides a versatile, highly-scalable
experimental setup, which may be used as a tool for understanding
quantum processes whose underlying physics can be somehow traced
to the concept of random walks, such as energy transport in
photosynthetic light-harvesting complexes
\cite{Mohseni2008,Plenio2008} and material band gap structures
\cite{Obuse2011}.

The structure of the article is as follows. In Sec. II, we
introduce the QRW model system, including dephasing, considered
here. The proposed experimental setup is described in Sec. III.
Numerical results are discussed in Sec. IV. Finally, we summarize our
results in the conclusion.

\section{A two-dimensional quantum random walk with dephasing}
A typical discrete quantum random walk comprises two operations: a coin-tossing operation
and a shift operation. Here, the coin-tossing operation is performed in the Hilbert space
$\mathcal{H}_p$ spanned by vectors $\{|H\rangle,|V\rangle\}$,
corresponding to the photon polarization. The random walk is
performed in the Hilbert space
$\mathcal{H}_X\otimes\mathcal{H}_Y$, corresponding to the
position of the photon in the transverse plane, spanned by vectors
$\{|i,j\rangle\}$ ($i,j$ integers), which indicate sites ($i,j$)
in the transverse plane ($i,j=...-2,-1,0,1,2...)$. The global
quantum system thus evolve in the Hilbert space
\begin{equation}
\mathbb{\mathcal{H}}=\mathcal{H}_{X}\otimes\mathcal{H}{}_{Y}\otimes\mathcal{H}_{p}.
\label{Eq:1}
\end{equation}

The state of the system is described by the density matrix
$\hat{\rho}^{(n)}$, which is transformed to a new density matrix
each step $n$ via the map
\begin{equation}
\hat{\rho}^{(n+1)}=\hat{P}^{(n)}\hat{S}_{Y}\hat{H} \hat{S}_{X}
\hat{H}\hat{\rho}^{(n)}\hat{H}^{\dagger} \hat{S}^{\dagger}_{X}
\hat{H}^{\dagger} \hat{S}^{\dagger}_{Y} \hat{P}^{(n)\dagger}.
\label{Eq:2}
\end{equation}
$\hat{H}$ denotes the Hadamard operator
\begin{equation}
\hat{H}=\frac{1}{\sqrt{2}}\left(\begin{array}{cc}
1 & 1\\
1 & -1
\end{array}\right),
\label{Eq:3}
\end{equation}
which acts on the polarization degree of freedom. The operators
$\hat{S}_{X}$ and $\hat{S}_{Y}$, which describe the walker's shift
in the transverse dimensions $x$ and $y$, independently, read as
\begin{equation}
\hat{S}_{X}=\sum_{i,j}|i-1,j,H\rangle\langle
i,j,H|+|i+1,j,V\rangle\langle i,j,V|, \label{Eq:4}
\end{equation}
and
\begin{equation}
\hat{S}_{Y}=\sum_{i,j}|i,j-1,H\rangle\langle
i,j,H|+|i,j+1,V\rangle\langle i,j,V|. \label{Eq:5}
\end{equation}

The coupling of the quantum walker with the environment is
described by pure dephasing \cite{kosik2006quantum}. The form of the unitary
dephasing operator considered here can be written as
\begin{equation}
\hat{P}^{(n)}=\sum_{ij}e^{-\frac{i}{2}\phi^{(n)}_{ij}\hat{\sigma}_z}|i,j\rangle\langle
i,j|,
\label{Eq:6}
\end{equation}
where $\phi^{(n)}_{ij}$ is a random phase matrix and
$\hat{\sigma_z}$ is the Pauli operator. Inspection of Eq.~(\ref{Eq:6}) shows that $\phi^{(n)}_{ij}$ represents a newly
introduced phase difference between the horizontal and vertical
polarizations at each site. Concerning this, we will consider
three physically relevant scenarios,
that can be easily implemented in the set-up proposed here (see Sec. III). In the
general case, the phase differences $\phi^{(n)}_{ij}$ are
independent random variables, but with the same probability
distribution. Moreover, the ensemble of phase differences
$\phi^{(n)}_{ij}$ can change from step $n$ to step $n+1$. In the following, we will
refer to this case as a QRW influenced by {\em dynamical spatial
disorder}.

\begin{figure}[t]
\includegraphics[width=9cm]{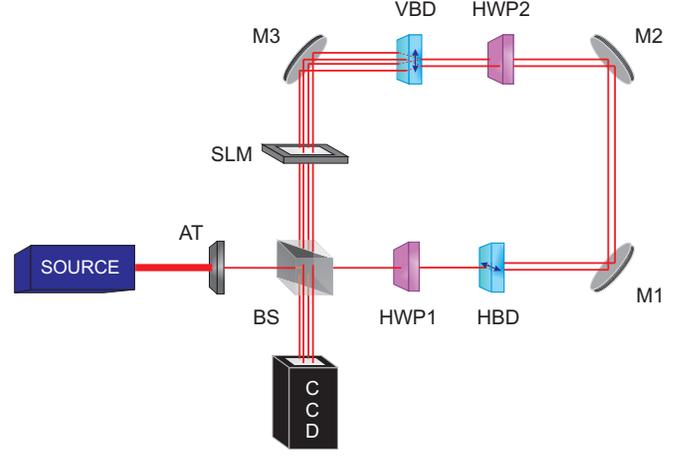}
\caption{(Color online) General scheme for the implementation of a
two-dimensional random walk with decoherence. AT: attenuator; BS:
beam splitter; HWP1 and HWP2: half-wave plates that act as coins
in the random walk; HBD and VBD: horizontal and vertical beam
displacers; M1,M2 and M3: mirrors; SLM: spatial light modulator;
CCD: spatial light sensor with single-photon sensitivity.}
\label{Fig:1}
\end{figure}

The easiest probability distribution that we can consider is an
uniform probability distribution. If phases can be chosen
arbitrarily between the extreme values $-\zeta$ and $\zeta$, there
is a constant probability $1/(2\zeta)$ to obtain any phase in this
interval. $\zeta=\pi$ is the maximal phase which we can have
between the two orthogonal polarizations. $\zeta=0$ corresponds to
the absence of any spatial disorder. If phases do not change
during propagation, even though they might differ from site to
site, i.e. $\phi^{(n)}_{ij}=\phi^{(n+1)}_{ij}$, then we have {\em
static spatial disorder}. Finally, if all phase differences are
the same for all sites, but they can still change from one step to
the following, we have {\em dynamical dephasing without spatial
disorder}.

The probability of detecting a photon in the site ($i,j$) is
\begin{equation}
p^{(n)}(i,j)=\langle i,j|Tr_{p}[\hat{\rho}^{(n)}]| i,j\rangle, \label{Eq:8}
\end{equation}
where the density matrix that describes the whole system is
traced out over the polarization degree of freedom $\left(Tr_{p}\right)$.

The spreading of the uncertainty of photon location is characterized by
dependence of the variance on the step index n

\begin{equation}
V^{(n)}=
\sum_{i,j}p^{(n)}(i,j)\left|\mathbf{r}_{ij}-\mathbf{\mu}\right|^{2}
. \label{Eq:9}
\end{equation}
where $\mathbf{r}_{ij}=(i,j)$ represents the lattice site with
indexes ($i,j$) and $\mathbf{\mu}$ is the mean position, i.e.,
$\mathbf{\mu}=\sum_{i,j}p^{(n)}(i,j)\mathbf{r}_{ij}$.

\begin{figure}[t]
\includegraphics[width=8cm]{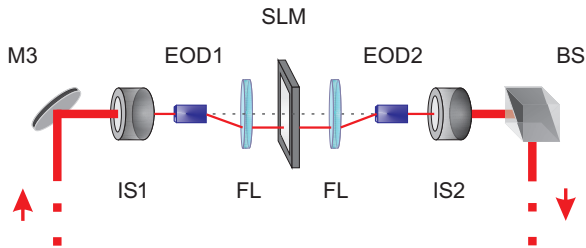}
\label{Fig:2} \caption{(Color online) Detail of the SLM part of
the setup allowing to generate dynamical spatial disorder. M3:
mirror; EOD1 and EOD2: electro-optic deflectors; FL: Fourier lens;
SLM: spatial light modulator; BS: beam splitter; IS1 and IS2:
imaging systems}
\end{figure}

\section{Experimental setup}
The main building block of the QRW setup is the multi-path
Mach-Zehnder-like configuration shown in Fig.~1. It allows to
make several runs of the QRW without the necessity of using a
large amount of optical components, as it is the case, for
instance, of the experiment described in \cite{Broome2010}. A
similar scheme, based on a single-path Mach-Zehnder-like
configuration has been used \cite{Schreiber2010,Schreiber2011}. However
in these cases, the walker moves in time, whereas in
our proposal the walker moves in the two-dimensional transverse
plane, offering a way to simplify the experimental implementation
of the two-dimensional QRW.

As source of photons one can use a highly attenuated short
coherent pulse, prepared by the combination of a photon source
and attenuator (AT), generating the initial state
\begin{equation}
|\Psi^{(0)}\rangle=|0,0\rangle\otimes\frac{1}{\sqrt{2}}\left(|H\rangle+i|V\rangle\right),
\label{Eq:10}
\end{equation}
where $(0,0)$ is the central site. The duration of the pulse has
to be sufficiently smaller than the time-of-fly through the setup
in one cycle. The transverse size of the Gaussian beam
profile of the pulse has to be carefully chosen, so that two
adjacent sites are not overlapping in the space due to
diffraction. For instance, by making use of a Gaussian beam of 2
mm beam waist, corresponding to a Rayleigh range of 23.6 m (for a
wavelength $\lambda$ = 532 nm), along with typical-sized optical
components, we could in principle perform a QRW of approximately
more than 20 steps. The number of steps can be further improved by
applying smaller beams together with a re-focusing system placed
along the walker's path. Alternatively, a spontaneous parametric
down-conversion source can be used, provided each down-converted
photon is generated in a pure state.  Then, the photon is
transmitted via the beam splitter (BS) to the system.

To get a clearer picture of the working of the quantum random
walk, let us consider in detail the quantum state of the photon in
its first passage through the system. First, the polarization
state of the photon is changed by the half-wave plate (HWP1) to
$\frac{1}{2}|0,0\rangle\otimes\left[(1+i)|H\rangle+(1-i)|V\rangle\right]$,
i.e., the Hadamard ($\hat{H}$) operation is applied. After this
transformation, the photon is displaced by the horizontal beam
displacer (HBD) along the $x$ axis according to its polarization,
as described by the shift operator $\hat{S}_{X}$. The photon is
now in the state
$\frac{1}{2}\left[(1+i)|-1,0,H\rangle+(1-i)|1,0,V\rangle\right].$
A second half-wave plate (HWP2) implements a new Hadamard
transformation, which transforms the quantum state to
$\frac{1}{\sqrt{8}}\{(1+i)\left(|-1,0,H\rangle+|-1,0,V\rangle\right)+(1-i)\left(|1,0,H\rangle-|1,0,V\rangle\right)\}$.
The vertical beam displacer (VBD) shifts the position of the
photon along the $y$ axis. After this, the quantum state of the
photon reads

\begin{eqnarray}
|\Psi^{(1)}\rangle &=&\frac{1}{\sqrt{8}}\left[(1+i)\left(|-1,-1,H\rangle+|-1,1,V\rangle\right)\right.\nonumber\\
& &+\left.(1-i)\left(|1,-1,H\rangle-|1,1,V\rangle\right)\right].
\label{Eq:11}
\end{eqnarray}
The spatial light modulator (SLM) is used to introduce random
phases, given by the phase-matrix $\phi^{(n)}_{ij}$. After the
photon passes through the SLM, the whole cycle is repeated with
high probability. If not, the photon escapes through the other
output port of the beam-splitter, which directs it towards a
highly sensitive (single-photon sensitivity) intensified
CCD camera with integrated photo-multipliers, which allows to
spatially resolve a weak signal with a high efficiency ($\sim$90\%)
\cite{camera}. Moreover, losses in the setup can be compensated by
an increase of the amplitude of pulse. The size of whole array of
beams can be reduced by an auxiliary imaging system in order to
fit on the limited size of sensitive area of CCD.

\begin{figure}[t]
\begin{center}

\begin{tabular}{c c}
(a)&\\
 &  \includegraphics[width=5.0 cm]{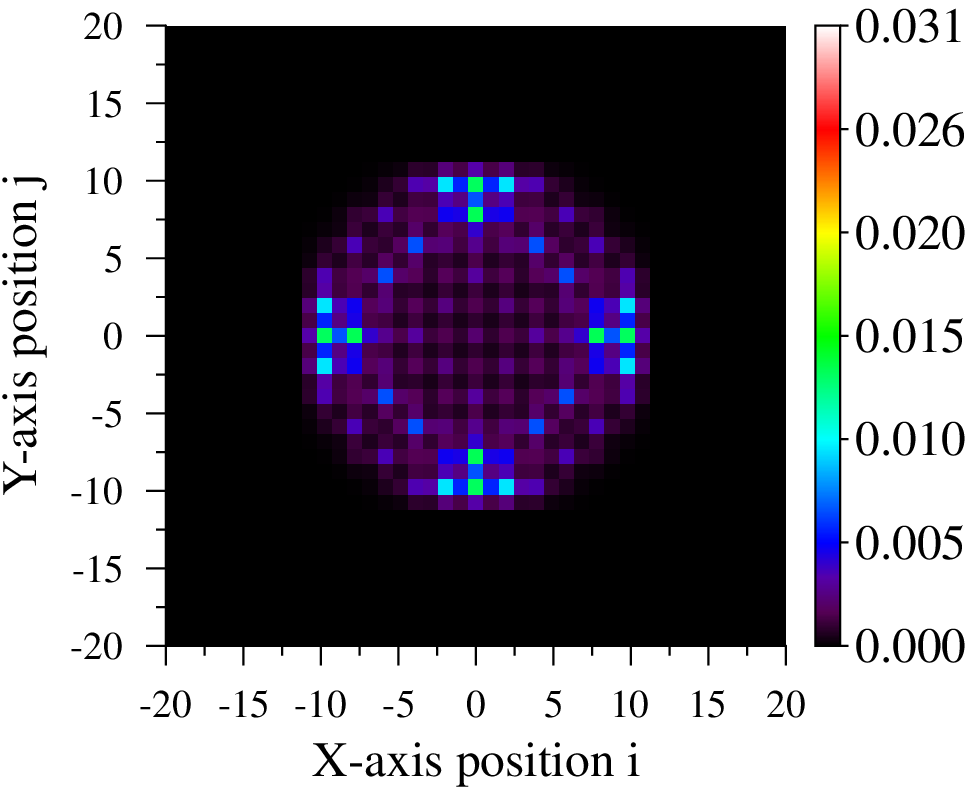}\\
(b) & \\
 &  \includegraphics[width=5.0 cm]{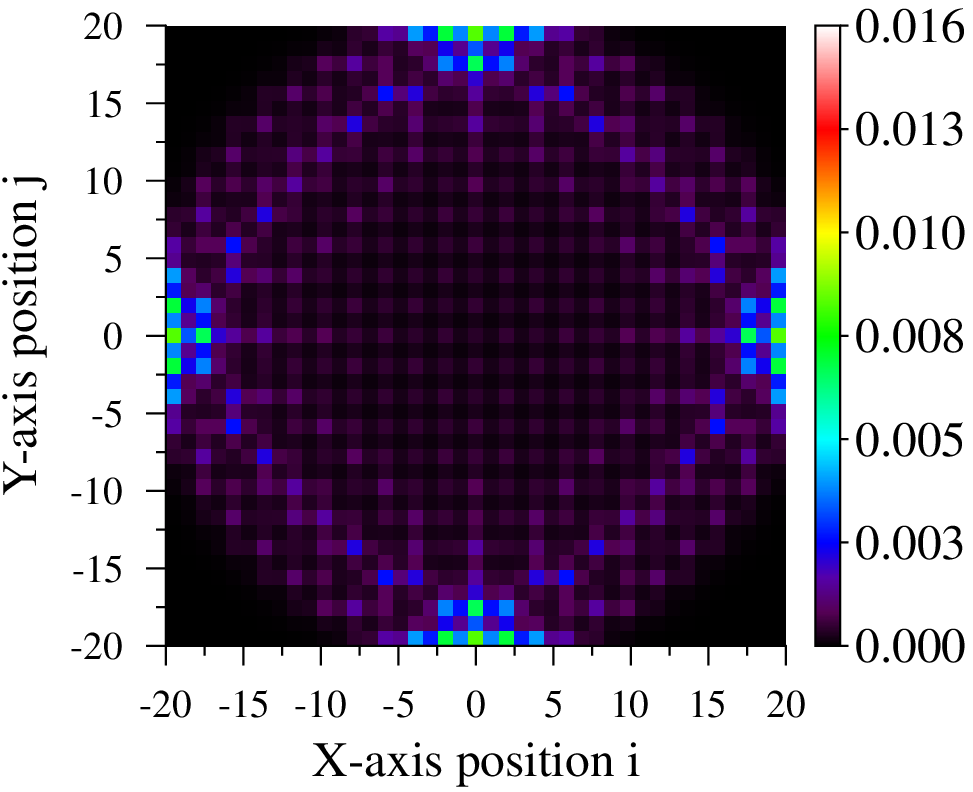}

\end{tabular}

\label{Fig:3}

\caption{(Color online) Probability distribution function $p^{(n)}(i,j)$, corresponding
to the position of the photon, for a two-dimensional quantum
random walk with no dephasing after a) $10$ steps and b) $20$
steps.}
\end{center}
\end{figure}

\begin{figure} [t]
\begin{center}
\begin{tabular}{c c}
(a)&\\
& \includegraphics[width=8 cm]{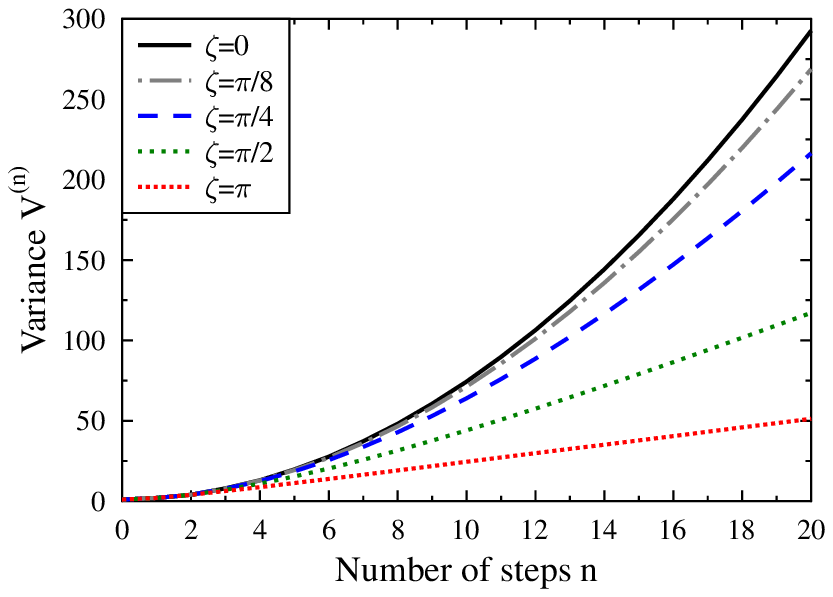}\\
(b) & \\
& \includegraphics[width=8 cm]{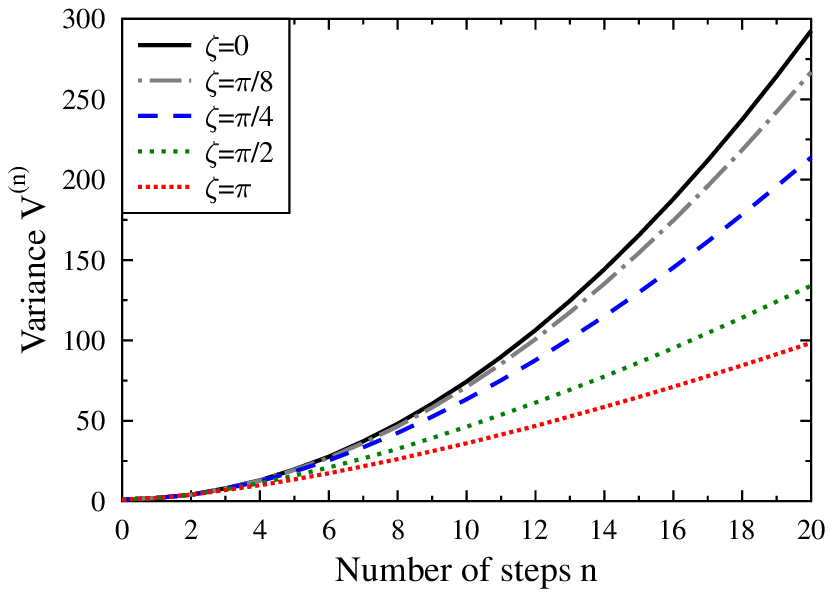}
\end{tabular}

\caption{(Color online) Spreading of the position of the photon ($V^{(n)}$)
as a function of the number of steps for several values of $\zeta$
and different types of dephasing. a) Dynamical spatial disorder.
b) Dynamic dephasing without spatial disorder. The results are
obtained averaging over 500 different realizations of the matrix
$\phi^{(n)}_{ij}$.}
\label{Fig:4}
\end{center}
\end{figure}

\subsection{Fast exchange of phase matrices}
A typical SLM  has a response time in the order of tens of
ms, which means that it is too slow for a fast phase-mask
exchange. For this reason, the transmission SLM shown in Fig.1 has
to be supplemented by additional components as it is shown in
Fig.2, which allows us to effectively generate the dynamical
spatial disorder and dynamical dephasing without spatial disorder
for a limited amount of steps. The time of exchange of phase
matrices can be done now in tens of ps
\cite{hisatake2005,hisatake2008}, which is three-orders faster
than it is indeed necessary in our experimental proposal, with a
typical time of flight of the pulse in the order of ns. The scheme
in Fig. 2 operates in the following manner. At the beginning, the
size of the whole array of beams is reduced by an imaging system
(IS1) to fit into the electro-optical deflector (EOD1). The EOD1,
together with a Fourier lens (FL), serve to address different
regions of the SLM along either the vertical or horizontal
direction, which realize then the random phase-matrices
$\phi^{(n)}_{ij}$ in all steps. Just changing the directions of
deflection of the array, both types of dynamical disorders can be
simulated. The EOD2 (and another FL) then serve to return all
beams back to the original direction of propagation, so that all
beams remain in the same position in the traverse plane at all
time.

\section{Numerical results}

\subsection{Quantum random walk}
Let us consider first the case when the SLM does not introduce any
phase shift ($\phi_{ij}^{(n)}=0$ for all $(i,j)$). This
corresponds to the case $\zeta=0$. Figure 2 shows the probability
distribution function $p^{(n)}(i,j)$ for (a) $n=10$ and (b) $n=20$ steps.
In both cases, the distribution shows a symmetrical shape
around the lines $x=0$ and $y=0$, with four groups of peaks
located along the $x$ and $y$ axes. The resulting symmetry comes
from the specific initial quantum state chosen in
Eq.~(\ref{Eq:10}). When the number of steps is increased, the
peaks move further away from the central site ($0,0$). The shapes
obtained in Fig.~3 correspond to the probability distributions of
a two-dimensional Grover walk
\cite{oliveira2006decoherence,Franco2011A}. The walker propagates
with ballistic speed, characterized by a quadratic dependence of the
variance $V^{(n)}$ with the step index, i.e., $V^{(n)}$
$\approx n^2$. This case is shown in Figs. 4(a) and 4(b),
corresponding to the case with $\zeta=0$.

\subsection{Quantum random walk affected by dephasing}
The dephasing effect introduced by the SLM allows to induce a transition from
the quantum to the classical random walk via two mechanisms.
First, as shown in Fig.~4(a), by means of dynamical spatial
disorder. The phase-matrix $\phi^{(n)}_{ij}$ shows independent and
randomly chosen values for each site, and it is refreshed each
step. The case $\zeta=0$ corresponds to the QRW with no dephasing. Increasing the
amount of disorder, characterized by a corresponding increase of
the parameter $\zeta$, reduce the spreading of $V^{(n)}$ as can be seen in Fig.~4(a).
In the  limiting case, which is reached for $\zeta=\pi$, the observed dependence ($\sim n$) of variance $V^{(n)}$ is a direct indication of the transition
to the classical regime of random walks.

The classical limit can also be reached by means of dynamical
dephasing without spatial disorder, as it is shown in Fig.~4(b).
The reduction of the uncertainty of

\begin{figure}[H]
\begin{center}
\begin{tabular}{c c}
(a) &\\
& \includegraphics[width=5 cm]{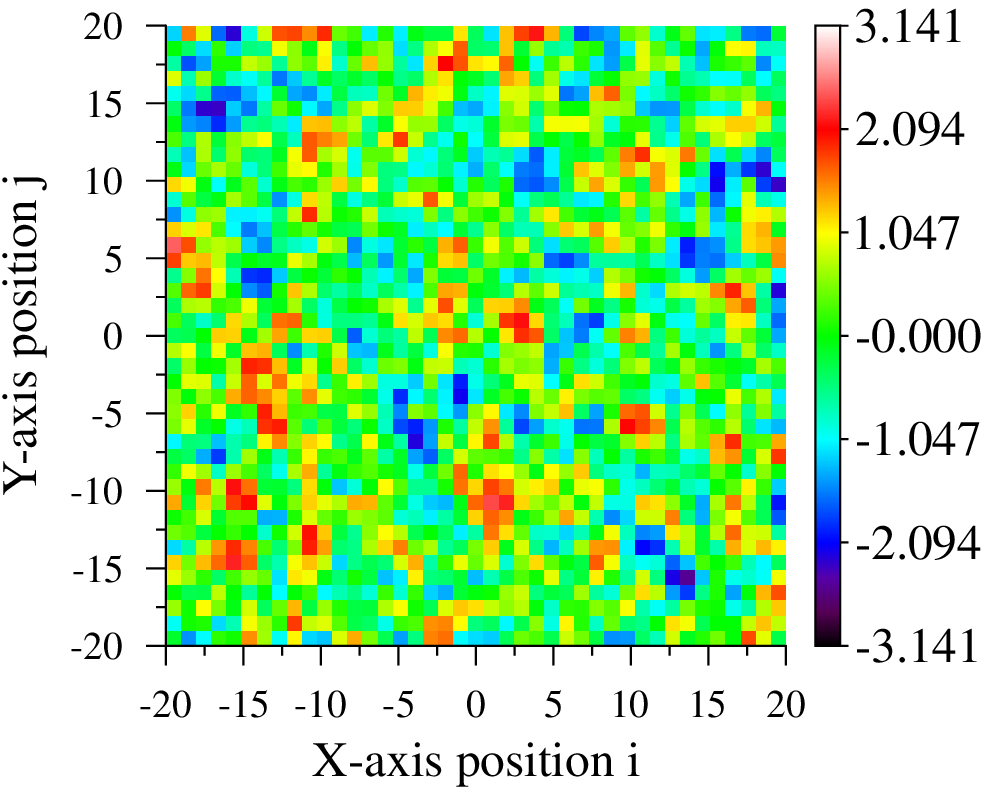}\\
(b) & \\
& \includegraphics[width=5 cm]{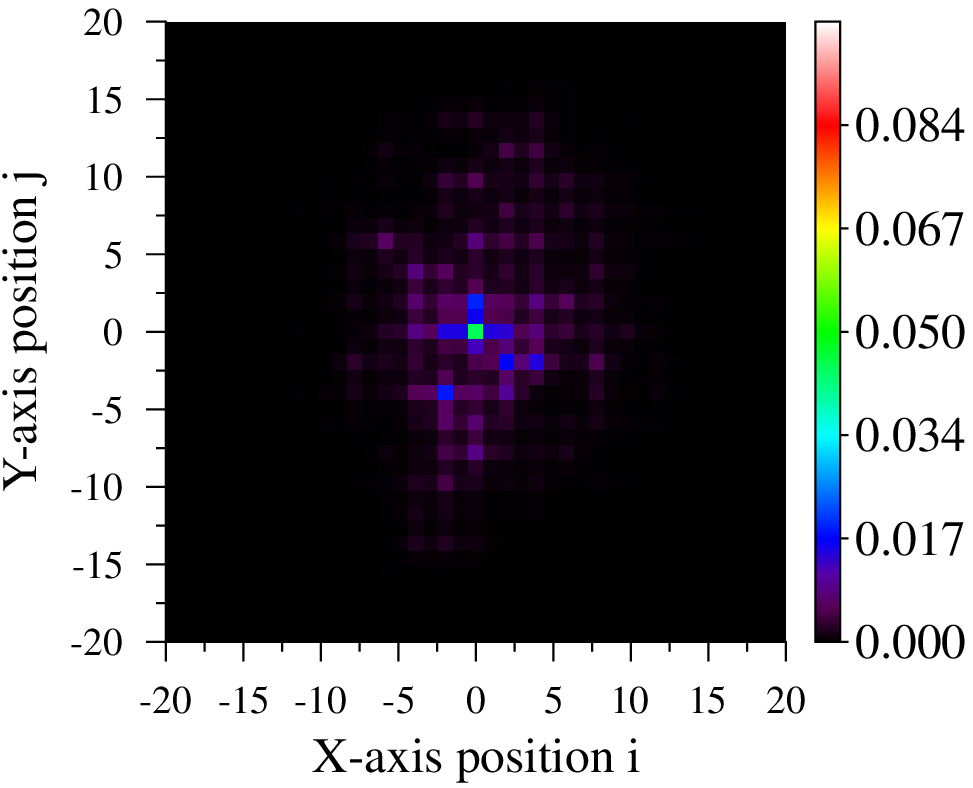}\\
(c) & \\
& \includegraphics[width=5 cm]{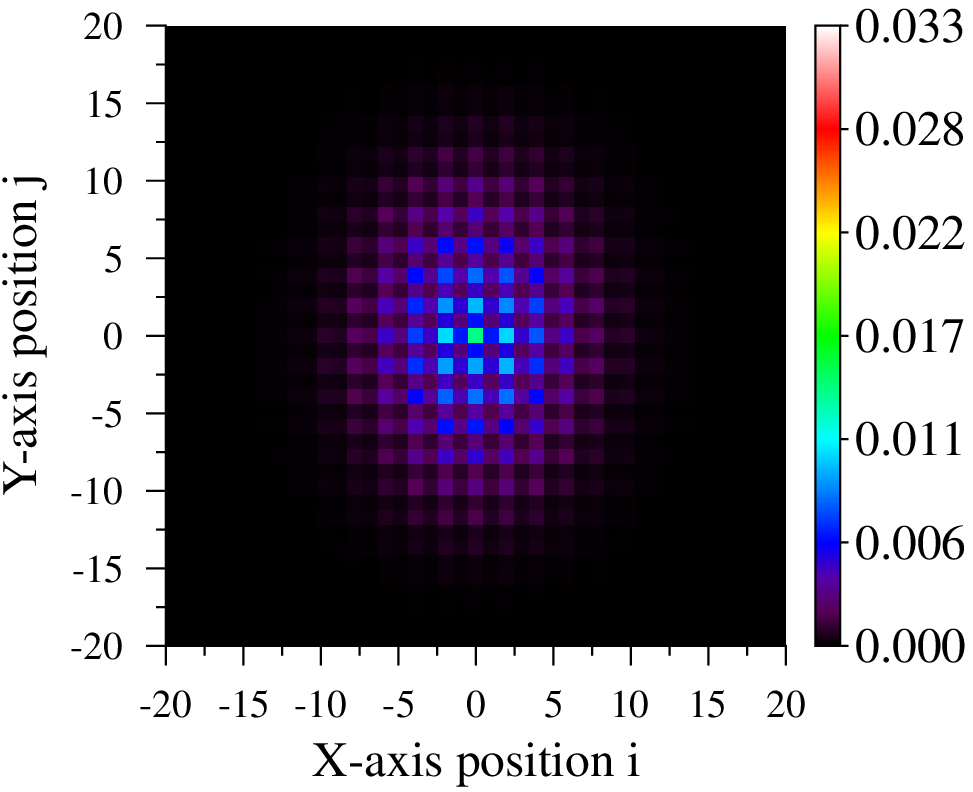}\\
(d) & \\
& \includegraphics[width=5 cm]{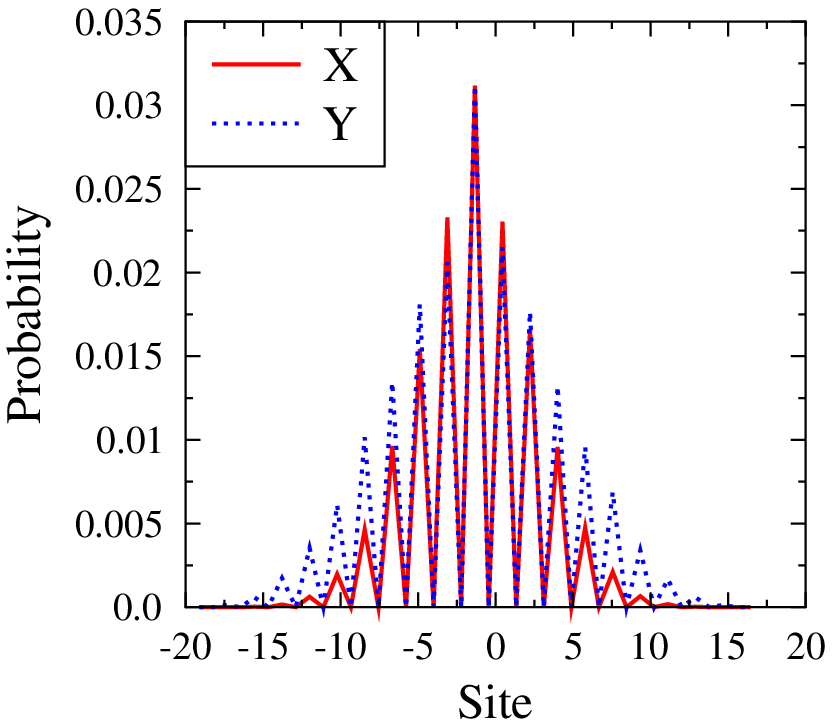}
\end{tabular}
\caption{(Color online) Observation of the spatial Anderson localization.
a) Example of a matrix $\phi^{(n)}_{ij}$ for $\zeta=\pi$ that
leads to the Anderson localization. b)
Corresponding probability distribution $p^{(n)}(i,j)$ after $20$ steps. c)
Averaged probability distribution over $500$ realizations. d) Cuts
of the data shown in Fig.~5(c) along the $X$ and $Y$ axes passing the site (0,0).}
\end{center}
\label{Fig:5}
\end{figure}

\noindent the photon position is less
dramatic than in the case with dynamical spatial disorder. For the
dynamical spatial disorder, $V^{(n)}\sim 51.25 $ for
$\zeta=\pi$ after 20 steps. On the contrary, for dynamical
dephasing without spatial disorder, we have  $V^{(n)} \sim
98.45$ under the same conditions. Indeed, the $n$-dependence
of the typical deviation, characteristic of the classical regime,
it is not yet reached after $20$ steps, as is readily observed in
Fig.~4(b).

\subsection{Anderson localization}
In the context of our discussion, the Anderson localization is the
reduction of spreading of the uncertainty of the photon position
\citep{anderson1958absence}. We will demonstrate that this effect
can also be observed in the set-up considered here. In
\cite{Schwartz2007}, Anderson localization was observed in the
transverse plane of a light beam passing through a crystal with
random static fluctuations of the index of refraction.
Since the randomness in the index of refraction
is affecting only the phase of the propagating beam, it is
possible to imitate these phase-fluctuations with a SLM, under
the consideration of static spatial disorder, since Anderson
localization does not appear with dynamical spatial disorder.

In Fig.~5(a) we present a typical profile of the phase-matrix
$\phi^{(n)}_{ij}$, independent of n, which leads to beam localization. Fig.~5(b) shows the
corresponding probability distribution of the photon position for
this specific phase profile. Notice that it
contains a strong peak located in the middle of the lattice. The
presence of Anderson localization is confirmed in Fig.~5(c), where
we show the averaged probability distribution function for
$\zeta=\pi$, exhibiting an exponential suppression of
probabilities for sites distant from the center. For the sake of
clarity, we also plotted in Fig.~5(d) two cuts of the averaged
probability distribution along the $X$ and $Y$ axes, to highlight
this feature.

\section{Conclusion}
We have put forward a new, highly scalable and easily implemented
experimental configuration to observe spatial two-dimensional random walks
under a great variety of circumstances, by means of the
implementation of two consecutive one-dimensional random walks.
The proposal makes use of only a small amount of simple optics components, and
allows us to simulate many different quantum systems and protocols
based on the quantum random walk concept. Additionally, by carefully
controlling the amount and type of disorder present in the system,
we have shown the effects of different environmental effects:
dynamical spatial disorder, dynamical dephasing without spatial
disorder and static spatial disorder. The last case drove us to
the observation of Anderson localization. The control of environmental
effects is paramount importance in nearly all quantum systems. In some cases,
it is even crucial to understand the dynamics experimentally observe.  For instance, in light
harvesting complexes \cite{Robentrost2009}, the interplay between coherent evolution and noise is a critical
ingredient, which allows to obtain a high efficiency energy transport.

\section*{Acknowledgement}
The authors thank to Dr. Jan Soubusta for useful discussions.
This work was supported by Project Nos. FIS2010-14831 and FET-Open
255914 (PHORBITECH).  J.Sv. thanks the project FI-DGR 2011 of The
Catalan Government. This work has also supported in part and by
projects COST OC 09026, CZ.1.05/2.1.00/03.0058 of the Ministry of
Education, Youth and Sports of the Czech Republic and by project
PrF-2012-003 of Palack\'{y} University.

\bibliographystyle{apsrev4-1}

\end{document}